\documentclass{webofc}

\usepackage[varg]{txfonts}
\usepackage{hyperref}
\usepackage{url}

\hypersetup{colorlinks=true,citecolor=blue,urlcolor=blue,linkcolor=blue}

\usepackage{xspace}
\usepackage{amssymb, amsmath}
\usepackage{subcaption}
\usepackage{lineno}
\usepackage[Symbol]{upgreek}
\usepackage{hyperref}

\newcommand{\pp}{pp\xspace}
\newcommand{\ppb}{p--Pb\xspace}

\newcommand{\pbpb}{Pb--Pb\xspace}

\newcommand{\etaphi}{\ensuremath{\eta - \varphi}\xspace}

\newcommand{\kt}{\ensuremath{k_{\mathrm{T}}}\xspace}
\newcommand{\pt}{\ensuremath{p_{\mathrm{T}}}\xspace}
\newcommand{\ptjet}{\ensuremath{p_{\mathrm{T, jet}}}\xspace}

\newcommand{\vz}{\ensuremath{\mathrm{V0}}\xspace}
\newcommand{\kzs}{\ensuremath{\mathrm{K^{0}_{S}}}\xspace}
\newcommand{\lz}{\ensuremath{\Lambda}\xspace}
\newcommand{\alz}{\ensuremath{\bar\Lambda}\xspace}

\begin{document}
\pagenumbering{arabic}

\title{Measurement of strange baryon production in charged-particle jets in pp and p-Pb collisions with ALICE}

\author{\firstname{Gijs} \lastname{van Weelden}\inst{1,2}\fnsep\thanks{\email{g.van.weelden@cern.ch}}, on behalf of the ALICE Collaboration
}

\institute{Nikhef, Science Park 105, 1098 XG Amsterdam
\and
           Utrecht University, Heidelberglaan 8, 3584 CS Utrecht
          }

\abstract{
The enhancement of the ratio of strange to non-strange hadrons with increasing charged-particle multiplicity density is often interpreted as a signature of the medium formed in heavy-ion collisions.
However, this effect has also been observed in smaller systems, such as \pp and \ppb collisions.
Understanding this effect requires a detailed description of the production mechanisms of strange hadrons in and out of jets.
In this manuscript published results on the production of $\mathrm{K^{0}_{S}}$, $\Lambda (\bar\Lambda)$, $\Xi^\pm$ and $\Omega^\pm$ hadrons in charged-particle jets and the underlying event in pp and p-Pb collisions measured by ALICE are presented.
In addition, a new analysis of in-jet fragmentation into $\Lambda$ and $\mathrm{K^{0}_{S}}$ hadrons in pp collisions at $\sqrt{s} = 13.6$ TeV from LHC Run 3 data is shown.
}

\maketitle

\section{Introduction}\label{sec:Introduction}
In heavy-ion collisions, a medium of deconfined quarks and gluons is formed.
One signature of this medium is the enhancement of the ratio of strange to non-strange hadrons with increasing charged-particle multiplicity density.
However, this effect has also been observed in smaller systems, such as \pp and \ppb collisions, where such a medium is not expected.

The yields of \kzs, \lz (\alz), $\Xi^\pm$, and $\Omega^\pm$ hadrons were measured as a function of charged-particle multiplicity in \pp, \ppb and \pbpb collisions by the ALICE Collaboration \cite{ALICE:2016fzo}.
The yields of these strange hadrons were found to increase with multiplicity at a rate faster than that of the pion yields.
Questions remain on the mechanisms driving the production of these hadrons and their increase with increasing charged-particle multiplicity.
In these proceedings, possible ways to disentangle the strangeness production mechanisms in jets and in the underlying event (UE) are investigated.
A published analysis \cite{ALICE:2022ecr} on the production of \kzs, \lz (\alz), $\Xi^\pm$, and $\Omega^\pm$ in charged-particle jets and in the UE in \pp and \ppb collisions is presented.
In particular, this is the first measurement of $\Omega^\pm$ production in jets.
Additionally, a new analysis is discussed on in-jet fragmentation into \kzs and \lz (\alz) using Run 3 ALICE data.

\section{Strange hadron yields in charged-particle jets}\label{sec:Yields}
Jets are reconstructed using the anti-\kt algorithm \cite{Cacciari:2011ma, Cacciari:2005hq}, with a jet resolution parameter $R = 0.4$, using tracks with transverse momentum $p_\mathrm{T, track} > 0.15$ GeV/$c$ and pseudorapidity $|\eta_\mathrm{track}| < 0.8$, across the full acceptance in azimuth $\varphi$.
The pseudorapidity of reconstructed jets is required to be in the range $|\eta_\mathrm{jet}| < 0.35$, ensuring the jets are fully contained within the acceptance region of reconstructed weak decays.
In addition, a transverse momentum cut of $\ptjet > 10$ GeV/$c$ is applied to ensure the jet originates from a hard scattering process.

This analysis considers the \vz particles (\kzs, \lz, and \alz), as well as the cascades (here meaning $\Xi^\pm$ and $\Omega^\pm$).
The following decay channels were searched for, with their corresponding branching ratios ($BR$):

\begin{align*}
	\kzs &\rightarrow \uppi^+ + \uppi^-
		&& BR = (69.20 \pm 0.05)\%, \\
	\lz(\alz) &\rightarrow \mathrm{p} (\bar{\mathrm{p}}) + \uppi^- (\uppi^+)
		&& BR = (63.9 \pm 0.5)\%, \\
	\Xi^- (\bar\Xi^+) &\rightarrow \lz (\alz) + \uppi^- (\uppi^+)
		&& BR = (99.887 \pm 0.035)\%, \\
	\Omega^- (\bar\Omega^+) &\rightarrow \lz (\alz) + \mathrm{K}^- (\mathrm{K}^+)
		&& BR = (67.8 \pm 0.7)\%.
\end{align*}

The \vz and cascade candidates are reconstructed based on the topology of their weak decays, within a pseudorapidity range of $|\eta| < 0.75$.
The purity is improved by identifying charged tracks based on their energy deposition in the Time Projection Chamber subdetector.
The invariant mass is used to extract the signal and combinatorial background.
In addition, candidates that satisfy the reconstruction criteria for multiple particles, are rejected.

Particles are defined as within the jet cone (JC) if their distance in the \etaphi plane with respect to the jet axis is less than the jet resolution parameter.
This includes both particles associated with the hard scattering and particles from the UE.
The UE contribution is estimated with the perpendicular cone method (PC) \cite{ALICE:2022ecr}, where the yields of strange hadrons are measured within a cone of size $R$, placed at the same $\eta$ as the jet and at an angle $\Delta \varphi = \pi/2$.

The \pt-differential yields $\mathrm{d}\rho/\mathrm{d}\pt$ of the particles are calculated by

\begin{equation}
	\frac{\mathrm{d} \rho}{\mathrm{d} \pt} =
		\frac{1}{N_\mathrm{ev}}
		\times
		\frac{1}{A_\mathrm{acc}}
		\times
		\frac{\mathrm{d} N}{\mathrm{d} \pt},
\end{equation}

where $\mathrm{d}N/\mathrm{d}\pt$ is the \pt-differential particle production yields, $N_\mathrm{ev}$ is the number of events, $A_\mathrm{acc}$ is the area of the \etaphi acceptance for a given selection.
The \pt-differential yields of in-jet particles can then be extracted by subtracting the PC selected yields from the JC selected yields.

\begin{figure*}[!ht]
\centering
\includegraphics[width=12cm,clip]{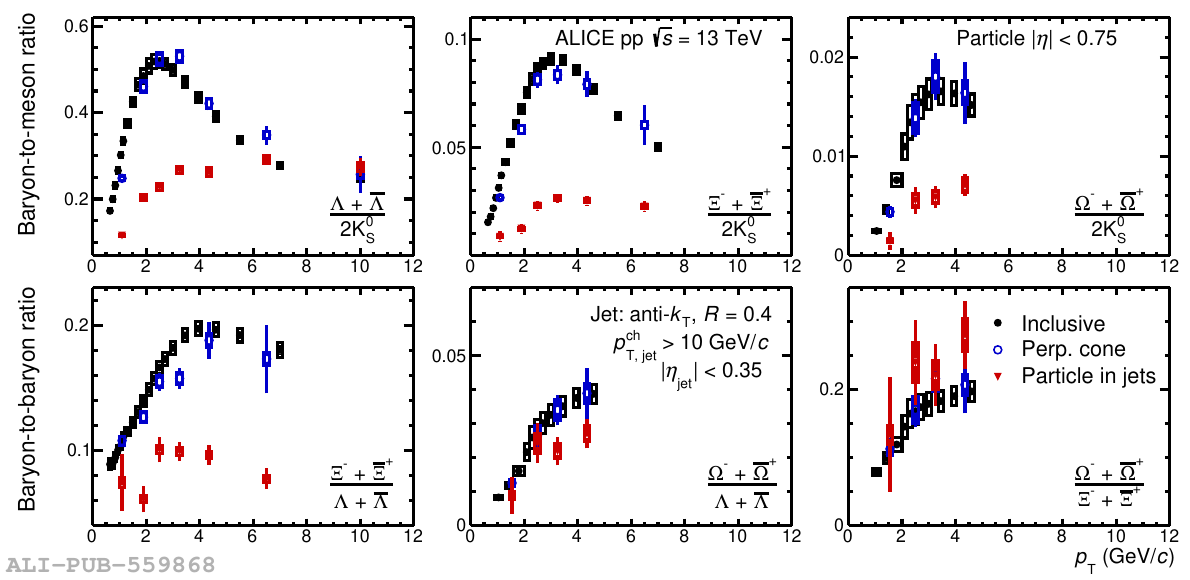}
\caption{\pt-dependent baryon-to-meson (top) and baryon-to-baryon (bottom) yield ratios for inclusive, in-jet, and PC selected particles in \pp collisions.}
\label{fig:YieldRatiospp}
\end{figure*}

Figure \ref{fig:YieldRatiospp} shows the baryon-to-meson and baryon-to-baryon ratios for the measured strange particles in \pp collisions.
For each of these ratios, the UE yields closely follow the inclusive yields.
For the baryon-to-meson ratios and $\Xi/\lz$, an enhancement is seen in the inclusive and UE yields with respect to the in-jet yields.
For the $\Omega/\lz$ and $\Omega/\Xi$ ratios, this enhancement is not seen, but the uncertainties are too large to make any definitive statements.

\begin{figure*}[!h]
\centering
\includegraphics[width=12cm,clip]{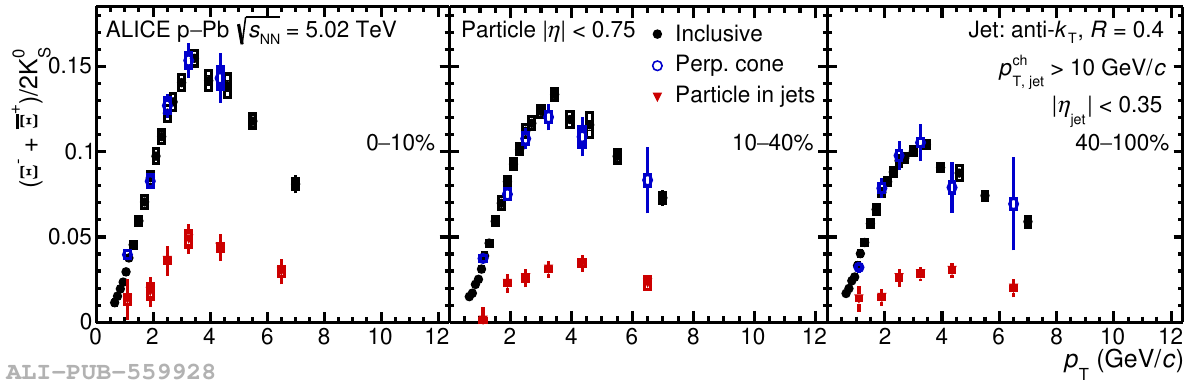}
\caption{$\Xi/\kzs$ yield ratios for inclusive, in-jet, and PC selected particles in \ppb collisions in different multiplicity classes.}
\label{fig:YieldRatiospPb}
\end{figure*}

Figure \ref{fig:YieldRatiospPb} shows the $\Xi/\kzs$ yield ratio in \ppb collisions for different multiplicity bins: high (0-10\%), intermediate (10-40\%) and low (40-100\%).
The UE yields show an enhancement with increasing multiplicity, whereas the in-jet particle yields are consistent within uncertainties across multiplicity.
A similar enhancement is found for the ratios of $\lz/\kzs$ and $\Xi/\lz$.
Due to the low number of $\Omega$ candidates in the studied data sample, the ratios of $\Omega/\kzs$, $\Omega/\lz$, and $\Omega/\Xi$ were only measured for Minimum Bias (MB) \ppb collisions.
The results are found to be consistent within uncertainties for MB \pp and MB \ppb collisions.

\section{Jet fragmentation}\label{sec:Fragmentation}
A different approach to studying the production of strange hadrons in jets is to measure the jet fragmentation functions.
The jet fragmentation function for a hadron is defined as the number density of hadrons in a jet, that carry a fraction $z$ of the longitudinal momentum of the jet.
The longitudinal momentum fraction is calculated by projecting the particle 3-momentum vector $\mathbf{p}_\mathrm{i}$ onto that of the jet ($\mathbf{p}_\mathrm{jet}$) and normalising to the jet momentum

\begin{equation}
	z_i = \frac{\mathbf{p}_\mathrm{i} \cdot \mathbf{p}_\mathrm{jet}}{\mathbf{p}_\mathrm{jet}^2}.
\end{equation}

Due to the relatively long lifetimes of \kzs (2.6844 cm/$c$) and \lz (7.89 cm/$c$) \cite{ParticleDataGroup:2024cfk}, their decay daughters are removed from the jet clustering input by removing tracks with a large distance of closest approach to the primary vertex.
Therefore, the \vz candidates are reconstructed separately and added to the jet clustering input, yielding a collection of charged-particle + \vz jets (ch+\vz jets).

\begin{figure*}[!ht]
\centering
\begin{subfigure}[l]{4cm}
\includegraphics[width=6cm,clip]{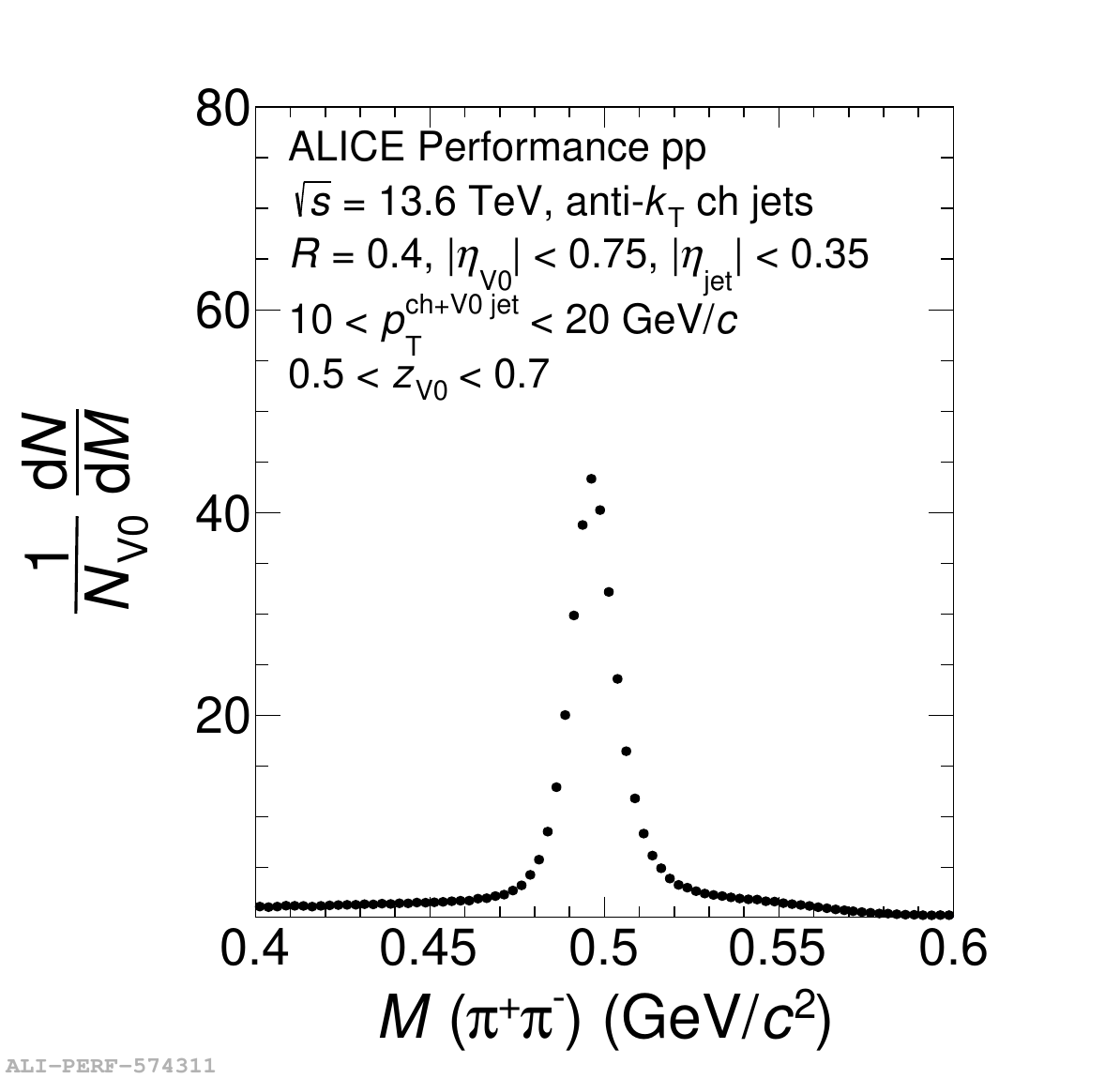}
\end{subfigure}
\hfill
\begin{subfigure}[l]{4cm}
\includegraphics[width=6cm,clip]{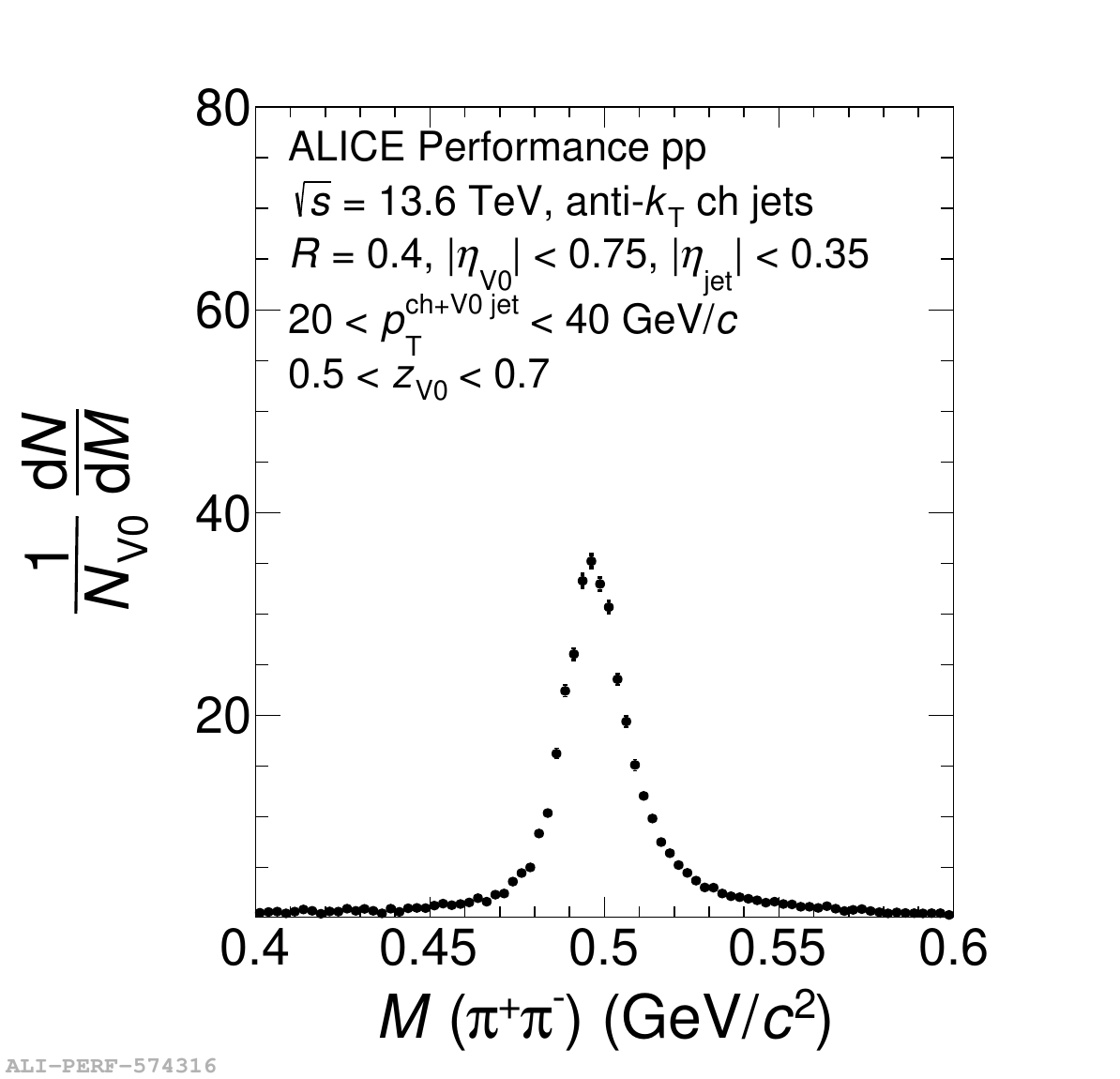}
\end{subfigure}
\hspace{2cm}
\caption{
Invariant mass spectrum of \vz candidates in ch+\vz jets around the \kzs mass, with $0.5 < z_{\vz} < 0.7$.
Left: $10 < \ptjet < 20$ GeV/$c$, right: $20 < \ptjet < 40$ GeV/$c$.
}\label{fig:k0sinjets}
\end{figure*}

To illustrate the performance of the ch+\vz jet approach in Run 3 ALICE data, figure \ref{fig:k0sinjets} shows the invariant mass spectra of \vz particles in ch+\vz jets around the \kzs invariant mass.
\vz particles with $0.5 < z < 0.7$ were selected in jets with $10 < \ptjet < 20$ GeV/$c$ (left) and $20 < \ptjet < 40$ GeV/$c$ (right).
The clearly visible invariant mass peaks with large numbers of candidates in the Run 3 data enables precise measurements of \vz fragmentation distributions in jets and study of the dependence on the jet momentum.

\section{Conclusions}\label{sec:Conclusions}
Studying the production of strange hadrons in jets can provide insight into the mechanisms that drive the increase of the strange to non-strange hadron ratio with charged-particle multiplicity density.
In these proceedings, the first measurement of $\Omega^\pm$ baryons in jets was presented, as well as results on \kzs, \lz (\alz), and $\Xi^\pm$ yields in jets, in \pp and \ppb collisions.
It is observed that the baryon-to-meson and the 2-strange-to-1-strange yield ratios show an enhancement with multiplicity in the UE, but not in jets.
Also, an approach to measuring in-jet fragmentation into \kzs and \lz (\alz) in \pp collisions using jets constructed from both charged particles and reconstructed \vz candidates was shown.
This approach showcases the possibility to identify \vz particles at high $z$ in Run 3 ALICE data.

\end{document}